\documentclass[a4paper,fleqn,usenatbib]{mnras}

\usepackage[T1]{fontenc}
\usepackage{ae,aecompl}

\usepackage{graphicx}	% Including figure files
\usepackage{amssymb}	% Extra maths symbols
\usepackage{savesym}
\usepackage{amsmath}
\savesymbol{iint}
\usepackage{txfonts}
\restoresymbol{TXF}{iint}

\title[LLAGNs at the centre of the Perseus cluster]{Discovery of five low luminosity active galactic nuclei at the centre of the Perseus cluster}

\author[Park et al.]{Songyoun Park,$^{1,2}$\thanks{E-mail: sypark@astro.snu.ac.kr}\thanks{Current address: Department of Physics and Astronomy, Seoul National University, Seoul 08826, Republic of Korea}
Jun Yang,$^{3,4}$
J. B. Raymond Oonk,$^{5,6}$
and Zsolt Paragi$^{4}$
\\
$^{1}$Department of Astronomy, Yonsei University, Seoul 03722, Republic of Korea \\
$^{2}$Korea Astronomy and Space Science Institute, Daejeon 34055, Republic of Korea \\
$^{3}$Department of Earth and Space Sciences, Chalmers University of Technology, Onsala Space Observatory, SE-439 92 Onsala, Sweden \\
$^{4}$Joint Institute for VLBI ERIC (JIVE), Postbus 2, 7990 AA Dwingeloo, the Netherlands \\
$^{5}$Netherlands Institute for Radio Astronomy (ASTRON), Postbus 2, 7990 AA Dwingeloo, the Netherlands \\
$^{6}$Leiden Observatory, Leiden University, PO Box 9513, NL-2300 RA Leiden, the Netherlands
}

\date{Accepted XXX. Received YYY; in original form ZZZ}

\pubyear{2016}

\begin{document}
\label{firstpage}
\pagerange{\pageref{firstpage}--\pageref{lastpage}}
\maketitle

\begin{abstract}
According to optical stellar kinematics observations, an over-massive black hole candidate has been reported by \citet{vandenBosch+12} in the normal early-type galaxy NGC~1277. This galaxy is located in the central region of the Perseus cluster. Westerbork Synthesis Radio Telescope (WSRT) observations have shown that NGC~1277 and other early-type galaxies in the neighbourhood have radio counterparts. These nuclear radio sources have stable flux densities on time scale of years. In order to investigate the origin of the radio emission from these normal galaxies, we selected five sources (NGC~1270, NGC~1272, NGC~1277, NGC~1278 and VZw~339) residing in the central 10 arcminute region of the Perseus cluster and requested to re-correlate the data of an existing very long baseline interferometry (VLBI) experiment at these new positions. With the re-correlation data provided by the European VLBI Network (EVN), we imaged the five sources with a resolution of about eight milliarcseconds and detected all of them with a confidence level above 5$\sigma$ at 1.4~GHz. They show compact structure and brightness temperatures above $10^7$~K, which implies that the radio emission is non-thermal. We rule out ongoing nuclear star formation and conclude that these VLBI-detected radio sources are parsec-scale jet activity associated with the supermassive black holes in low luminosity active galactic nuclei (LLAGNs), although there are no clear signs of nuclear activity observed in the optical and infrared bands. Using the fundamental plane relation in black holes, we find no significant evidence for or against an extremely massive black hole hiding in NGC~1277.  
\end{abstract}


%%%%%%%%%%%%%%%%%%%%%%%%%%%%%%%%%%%%%%%%%%%%%%%%%%

%%%%%%%%%%%%%%%%% BODY OF PAPER %%%%%%%%%%%%%%%%%%

\section{Introduction}

Recently, the early-type galaxy NGC~1277 in the Perseus cluster received much attention because, according to stellar kinematics observations, it most likely hides an over-massive black hole of $\sim$$10^{10}~\mathrm{M}_{\sun}$ \citep{vandenBosch+12, Yildirim+15}. Galaxies with the most massive black holes are interesting on their own, because they may serve as probes for black hole formation models \citep[][and references therein]{Khan+15}, may indicate capture of a runaway black hole \citep{ShieldsBonning13}, and may constrain black hole--galaxy co-evolution models \citep{Croton+06,Somerville+08}. 

The most massive black holes are also excellent targets to test accretion theory models \citep{Fabian+13}. The ultra-massive black hole in NGC~1277 may be still actively accreting mass although it is located in a normal galaxy according to optical and infrared observations. Westerbork Synthesis Radio Telescope (WSRT) observations revealed a compact radio counterpart with a total flux density of 2.9~mJy at 1.4~GHz \citep[e.g.][]{Sijbring93}. We also noticed that there are more radio detections of the normal early-type galaxies in the wide-field WSRT image of the Perseus cluster. Their radio emission is likely related to parsec-scale jet activity, powered by their central supermassive black holes, rather than star formation. In order to verify this scenario, we imaged five galaxies including NGC~1277 in the central region of the Perseus cluster with the very long baseline interferometry (VLBI) technique.

VLBI imaging observations are a direct and reliable way of revealing low-luminosity active galactic nuclei \citep[LLAGNs; $L_{\mathrm{H}\alpha}<10^{40}$~erg~s$^{-1}$,][]{Ho+97}. Supporting VLBI evidence for AGN emission often include quite high brightness temperatures ($T_{\rm B}\gg10^{5}$~K), radio luminosities exceeding that of nearby known nuclear supernova remnant (SNR) complexes ($L_{1.4 \rm ~GHz} > 10^{22} \rm ~W~Hz^{-1}$), and non-thermal flat radio spectra \citep[see][and references therein]{Alexandroff+12}. The radio emission may also come from nuclear star formation activity in the host galaxies \citep{Kimball+11, Padovani+11, Bonzini+13}, while most of the star formation-related radio emission typically has $T_\mathrm{B}\leq10^{4}$~K \citep{Condon+91}. Young radio supernovae (SNe) and SNRs may appear as compact sources in some cases \citep[e.g.][]{Kewley+00}, while they have a decaying flux density and are often resolvable at the later stage with milliarcsecond (mas) resolution \citep[e.g.][]{Perez-Torres+09}.   

The five early-type galaxies with WSRT radio counterparts are listed in Table~\ref{table1}. They were selected because they are within a circle of about 5 arcminutes from 3C~84, one of the central galaxies in the Perseus cluster. The selection criteria enabled us to re-use an existing European VLBI network (EVN) experiment, which was designed to study HI gas in NGC~1275/3C~84 (Oonk et al., in prep.), by re-correlation of the raw data at new positions. To get sub-mJy image sensitivity, we restrict our targets within the main beam of the Effelsberg 100~m radio telescope (full width at half maximum: about 10 arcminutes at 1.4~GHz). This instrumental restriction implies that we can only investigate a handful of the most centrally located early type galaxies in the Perseus cluster, and as such this sample is not complete in any sense.  

This paper is organized as follows. In Section 2, we describe the EVN observation and our data reduction methods. In Section 3, we present the VLBI detection of these WSRT sources. In Section 4, we discuss the possible explanations for the VLBI imaging results and the implications for an over-massive black hole in NGC~1277. In Section 5, we conclude that all the five early-type galaxies are most likely LLAGNs. 

Throughout this paper, we assume a cosmology with $\Omega_{m} = 0.3$, $\Omega_{\Lambda} = 0.7$, and $\rm H_{0}\,=\,70\,km\,s^{-1}\,Mpc^{-1}$. At $z=0.017$, the mean redshift of the five galaxies, an angular size of 1~mas corresponds to a linear size of 0.3~pc.

\begin{table*}
	\centering
	\caption{Summary of the EVN imaging results of the five early-type galaxies in the central region of the Perseus cluster. The columns are the following: source name, right ascension (J2000), declination (J2000), redshift, peak brightness, rms noise, restoring beam major and minor axes, position angle, model-fitted circular Gaussian component flux density, radius and brightness temperature lower limit.} 
	\label{table1}
	\begin{tabular}{@{}ccccc c r@{}l cccc r@{}l @{}}
		\hline
		Name & RA & Dec & z & Peak & RMS & \multicolumn{2}{c}{Major} & Minor & PA &  $S_{\nu}$  & Radius & \multicolumn{2}{c}{$T_{\rm{B}}$}  \\
		& (h m s) & ($\degr$ $\arcmin$ $\arcsec$) & & \multicolumn{2}{c}{(mJy~beam$^{-1}$)} & \multicolumn{2}{c}{(mas)} & (mas) & (\degr)& (mJy) &  (mas) & \multicolumn{2}{c}{($10^8\,\rm K$)} \\
		\hline
		NGC~1270 & 03:18:38.109 & +41:28:12.424 & 0.017 & 4.68 & 0.29 & ~10&.61 & 6.80 & $-$3.80 & 6.28 & 4.9 & $\geq$1&.7 \\
		NGC~1272 & 03:19:21.283 & +41:29:26.594 & 0.013 & 0.34 & 0.06 & ~10&.58 & 6.44 & $-$4.09 & 0.33 & 2.7 & $\geq$0&.3 \\
		NGC~1277 & 03:19:51.485 & +41:34:24.867 & 0.017 & 0.84 & 0.06 & ~10&.16 & 6.34 & $-$4.55 & 0.84 & 0.6 & $\geq$14&.8 \\
		NGC~1278 & 03:19:54.131 & +41:33:48.246 & 0.020 & 2.62 & 0.14 &  ~9&.33  & 5.70 & $-$5.58 & 2.97 & 2.3 & $\geq$3&.5 \\
		VZw~339   & 03:20:00.880 & +41:33:13.597 & 0.017 & 1.71 & 0.10 & ~10&.31 & 6.18 & $-$5.75 & 2.21  & 4.1 & $\geq$0&.8 \\
		\hline
	\end{tabular}
\end{table*}

\begin{figure*}
	\begin{center}
		\includegraphics[width=0.99\textwidth]{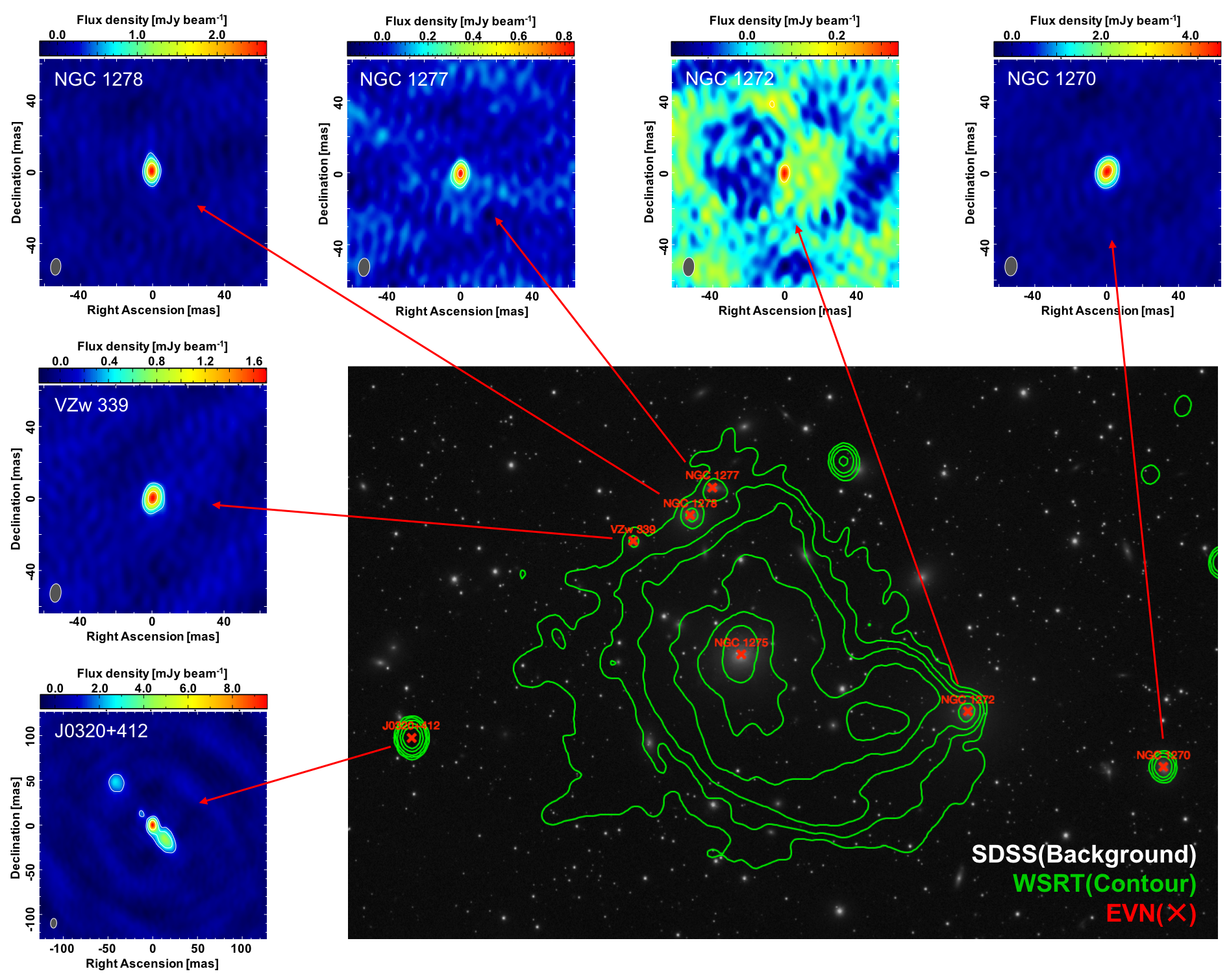}
		\caption{Combined image of the radio-optical images with compact radio sources detected from the EVN observation on milliarcsecond scale. The EVN maps of our five early-type galaxies and the phase-reference check source (J0320+412) are presented on the top and left-hand side. The contour levels on the VLBI images are powers of two times the 3$\sigma$ rms noise level. The green contours represent the 1.4~GHz WSRT radio continuum \citep{Sijbring93} which is overlaid on a \textit {SDSS} g band image.} 
		\label{figure1}
	\end{center}
\end{figure*}

\section[]{Observations and data reduction} 
Besides the five target galaxies listed in Table~\ref{table1}, we included J0320+412, a $\sim$50~mJy radio source that is located in our field of interest while not associated with the Perseus cluster, in the re-correlation. This check source was used to verify the quality of the data correlation and phase-referencing calibration. 

The European VLBI Network (EVN) observations were carried out at 1.4~GHz for 12 hours on 2012 October 18 (project code: EO009). The participating telescopes were Effelsberg (Ef), Westerbork Synthesis Radio Telescope (WSRT), Jodrell Bank (MK2), Onsala, Medicina, Noto, Torun, Zelenchukskaya, Badary and Urumqi. The data were recorded with a bit rate of 256$\,$Mbps using $2\times16$~MHz bandwidth and 2-bit sampling. The re-correlation was done in a single pass with the multi-phase centre correlation function provided by the EVN Software Correlator at JIVE \citep[SFXC;][]{Keimpema+15}. The re-correlation data have 2 seconds integration time and 32 channels for each sub-band. 

We used the Astronomical Image Processing System \citep[AIPS;][]{Greisen03} for the data reduction and followed standard procedures outlined in the EVN Data Analysis Guide\footnote{http:/$\!$/www.evlbi.org/user$\_$guide/guide/userguide.html}. A priori amplitude calibration was done with the measured gain curves and system temperatures for most of the telescopes except for Zelenchukskaya and Badary. The absence of amplitude calibration data at Russian stations will not significantly affect the final total flux density measurements since these measurements are mainly dependent on the short European baselines. Note that the WSRT phased-array data were flagged out because its synthesized beam is too narrow (about 15 arcsec) to include any target sources. We carried out fringe-fitting, bandpass calibration and self calibration on the in-beam calibrator 3C~84. After each step, we copied the calibrator solutions to each target source data set. The antenna primary beam correction was done using a simple circular Gaussian model \citep[e.g.][]{Cao+14}. The residual systematic amplitude calibration error is expected to be about 5 percent (1$\sigma$).  

The final editing, model-fitting of the $uv$-data and imaging were carried out in DIFMAP \citep{Shepherd+94}. Some data on the shortest baselines were also flagged out due to interference from the nearby very bright source 3C~84. We fitted a single circular Gaussian component to the observed visibility data for each of the target sources in order to constrain its size. Because of the low signal-to-noise ratios, phase self-calibration was avoided. To maximise the image sensitivity, we used natural weighing in the final clean images. The total flux density, the source positions (relative to 3C~84) and the angular size, measured via model-fitting, are listed in Table~\ref{table1}. The total flux density errors assume 5\% systematic error, in addition to the rms noise in the image. The estimated positional error is about 1$\,$mas. We also tried to fit a point source model to the visibility data. The fitting results are also as good as the ones observed by using a circular Gaussian model. Therefore, all the angular sizes are most likely upper limits.  

\section[]{Parsec-scale radio sources in the normal early-type galaxies} 
The EVN imaging results of the five early-type galaxies are shown in Figure~\ref{figure1}. All the target sources clearly reveal compact radio emission above the 5$\sigma$ detection level in the EVN images. In the dirty images, these sources clearly showed the beam structure, further supporting that all of them are genuine detections. These VLBI-detected radio sources are spatially coincident with the nuclei of these galaxies. Comparing with the position of the X-ray counterparts observed in NGC~1277, NGC~1278 and VZw~339 with \textit {Chandra} or \textit {ROSAT}, the position differences are within $\sim$0.1 arcsec, consistent with their formal position errors. 

The bright phase-referencing check source J0320+412 was treated in the same way as these target sources during the data reduction. It has a typical linear jet structure that is consistent with the results of using phase self-calibration. This again assured us that our in-beam phase-referencing calibration scheme worked well. 

Brightness temperatures of the sources were determined using the formula
\begin{equation}
	T_{\rm B}\, [\rm K]\,=\,1.22\,\times\,10^{12}(1\,+\,z)\frac{S_{\nu}}{\theta^{2}\nu^{2}} \,
\end{equation}
where $z$ is the redshift, $S_{\nu}$ is the flux density in Jy, $\theta$ is the radius of the fitted Gaussian model component in mas and $\nu$ is the observing frequency in GHz. Considering our angular sizes are an upper limit, the equation will provide an lower limit on the brightness temperature. The estimated brightness temperatures are $\geq3\times10^{7}$~K for the five sources. They are significantly larger than the typical value of $\sim$$10^{4-5}$~K, observed in star-forming regions. Therefore, the VLBI-detected radio emission most likely comes from non-thermal synchrotron radiation mechanism.

We also re-analysed 1.4~GHz Westerbork Synthesis Radio Telescope (WSRT) images observed in 1987, 1994 and 2003 \citep[G. de Bruyn, priv. comm.]{Sijbring93}. These images have an angular resolution of about 15~arcsec. Comparing the fluxes at these different epochs showed that there was no significant variability over 16 years for the five sources. 

The WSRT flux densities at 1.4~GHz are listed in Table~\ref{table2}. Comparing them with the EVN images, we found that more than 30 percent of radio emission was recovered in four out of five galaxies. In case of VZw~339, the EVN flux density was almost the same as that of WSRT. This implies that the radio emission of this galaxy mainly comes from a region quite compact on the parsec scale. The missing flux in the EVN maps may imply the existence of radio structure with an angular size smaller than the WSRT beam ($\sim$15~arcsec) while significantly larger than the EVN beam ($\sim$8~mas). This is a well-known instrumental effect caused by the absence of short baselines of the EVN observations. In the case of NGC~1272, the EVN only restored 3 percent of total flux density because it has significant jet emission on the kilo-parsec scale \citep{McBrideMcCourt14}.

\section{Discussion}

\subsection{Nature of the VLBI-detected radio emission}
Nuclear radio emission is ubiquitous in the most massive early-type galaxies \citep[e.g.][]{Brown+11}.
They are either related to nuclear stellar activities, such as star formation and supernova remnants, or jets powered by the central supermassive black holes.

It has been known for some time that many nearby elliptical and S0 galaxies contain nuclear radio sources smaller than about a few hundred parsecs, as shown by high resolution observations with the Very Large Array
(VLA) \citep{Sadler+89,Wrobel+91,Nyland+16}. Optical spectroscopy of these VLA-unresolved radio sources in nearby early-type galaxies showed that they are most likely powered by LLAGNs as opposed to star
formation, since they showed signs of non-thermal emission at a level that could not be explained by supernova remnants \citep{Sadler+89}. In most cases the spectra are characteristic of low-ionization nuclear emission line regions (LINER) and Seyfert galaxies \citep{Ho99}. Furthermore, when looked at with VLBI, a significant fraction had maximum brightness temperatures well exceeding $10^{5}$~K, often in flat/inverted spectrum compact cores, and/or had resolved jets \citep{Slee+94,Nagar+00,Nagar+02,Nagar+05}.

In the literature, it is common to refer to sources unresolved or barely resolved on arcsecond scales as compact radio sources. In what follows we will reserve the term compact for sources that have a significant fraction of their emission in milliarcsecond-scale structure, or are unresolved with VLBI.

\subsubsection{Ruling out stellar activity}
%NOTE: L(H-alpha, Sakai+2012) implies SFR(H-alpha)<0.3 M(sun)/yr for N1270,1272,1277
To test the star formation scenario, we estimated the total star formation rates using multi-wavelength data: radio, far-infrared (FIR), optical, and X-ray band. The FIR flux densities were extracted from \textit {IRAS}. The H$\alpha$ flux densities and X-ray luminosities were taken from the literatures \citep{Sakai+12,Santra+07}.
Total star formation rates, as obtained using the relation from \cite{Bell+03} for nuclear radio flux densities measured by the EVN, from \cite{Kennicutt+98} for the \textit {IRAS} FIR and the H$\alpha$ flux densities, and from \cite{Ranalli+03} for the X-ray luminosities, are consistent with normal star-forming galaxies (see Table~\ref{table2}). We note that the star formation rates obtained from these different wavelengths do not all agree. In particular the radio derived star formation rates are higher in most cases. This is another indication that the observed emission is likely not due to star formation only.

However, considering the very small physical area sampled by the EVN, the emission seen
on mas scales is not consistent with star formation in there early-type galaxies because the associated star formation rate per unit area is too high, $\sim$$10^{5-7}\,\rm{M_{\sun}\,yr^{-1}\,kpc^{-2}}$. This is comparable to some of the most extreme sites of star formation, for example in the 
luminous infrared galaxy NGC~4418, where star formation rates per unit area of $\sim$$10^{4.5}-10^{5.5}\,\rm{M_{\sun}\,yr^{-1}\,kpc^{-2}}$ were measured with the EVN \citep{Varenius+14}. However, none of the other typical tracers, such as strong optical emission lines and blue broadband colours, associated with intense star formation are 
observed in these objects (see also Sect.~\ref{llagn}). We therefore conclude that the scenario of
ongoing nuclear star formation can be ruled out.

Alternatively, the observed radio emission can be associated with young radio SNe or SNRs. 
The radio luminosity of our targets is comparable to that of the brightest radio supernovae in
nearby starburst galaxies with a radio luminosity of $\sim$$10^{20-22}\,\rm{W\,Hz^{-1}}$ 
\citep{Weiler+02}. A single or multiple radio SNe are required to explain the observed 
radio emission in our sources. However, young radio SNe do not seem to be responsible 
for the radio emission because no significant flux density changes were found in the 
WSRT observations for more than a decade.

\begin{table*}
	\centering
	\caption{The columns are the following: source name, 1.4~GHz flux density from WSRT \citep{Sijbring93}, 1.4~GHz flux density ratio of EVN (our observation) and WSRT \citep{Sijbring93}, implied star formation rates per unit area from our EVN observation, the total star formation rates derived from the radio (EVN), FIR, ${\rm H\alpha}$, and X-ray luminosities, and the X-ray luminosity measured by \textit {Chandra} \citep{Santra+07}. }
	\begin{tabular}{@{}c r@{}l c r@{}l c c r@{}l cc @{}}
		\hline
		Name &  \multicolumn{2}{c}{$S_{\nu}$} & Ratio & \multicolumn{2}{c}{SSFR$_{\rm radio}$} & SFR$_{\rm radio}$ & SFR$_{\rm FIR}$ & \multicolumn{2}{c}{SFR$_{\rm H\alpha}$} & SFR$_{\rm X-ray}$ & $\rm log\,\textit L_{\rm X}$  \\
		& \multicolumn{2}{c}{(WSRT)} & (EVN/WSRT) & \multicolumn{2}{c}{(EVN)} & (EVN) & & & & & (0.5-7 keV) \\
		& \multicolumn{2}{c}{(mJy)} &  & \multicolumn{2}{c}{($10^{6}~ \rm M_{\sun}\,yr^{-1}\,kpc^{-2})$} & ($\rm M_{\sun}\,yr^{-1}) $ & $\rm (M_{\sun}\,yr^{-1})$ & \multicolumn{2}{c}{$\rm (M_{\sun}\,yr^{-1})$} & $\rm (M_{\sun}\,yr^{-1})$ & (erg s$^{-1}$) \\
		\hline
		NGC~1270 &  ~~13&.10 & 0.48 & ~~~~~~~~~~~1&.0 & 2.6 & $<$0.6 & ~~~$<$0&.1 & -   & -     \\
		NGC~1272 &  ~~13&.00 & 0.03 & ~~~~~~~~~~~0&.4 & 0.2 & $<$0.3 & ~~~$<$0&.1 & -   & -     \\
		NGC~1277 &  ~~2&.85   & 0.29 & ~~~~~~~~~~~14&.5 & 0.6 & $<$0.8 & ~~~$<$0&.1 & 4.7 & 40.37 \\
		NGC~1278 &  ~~4&.90   & 0.61 & ~~~~~~~~~~~2&.2 & 1.9 & $<$0.8 & ~~~0&.9      & 0.9 & 39.65 \\
		VZw~339    &  ~~2&.25   & 0.98 & ~~~~~~~~~~~0&.1 & 1.2 & -          & ~~~0&.2      & 0.6 & 39.49 \\
		\hline
	\end{tabular}
	\label{table2}
\end{table*}

\subsubsection{Evidence for the LLAGN scenario}\label{llagn}
Flat or inverted spectrum radio cores in quasars are well-known to be related to the optically thick base of synchrotron jets \citep{BlandfordKonigl79}. Partially synchrotron self-absorbed jets are also ubiquitous in LLAGNs, but in that case these are related to radiatively inefficient accretion in ``hard state'' black holes \citep[e.g.][]{Nagar+05}. There are however 
alternative explanations for compact cores in LLAGN, like free-free emission from an X-ray heated accretion disk 
wind and at relatively lower brightness temperature \citep[e.g.][]{BondiPerezTorres10}, or synchrotron emission absorbed by the surrounding thermal gas in the nuclear region 
\citep[free-free absorption,][]{LalHo10,Varenius+14}. 

Low resolution WSRT observations at 1.4 and 5~GHz by \citet{Sijbring93} show that three of our target sources (NGC~1270, NGC~1277 and NGC~1278) have flat spectra ($\alpha_{5/1.4}>-0.5$, where $S_{\nu}\propto\nu^{\alpha}$). These flat spectra, in addition to the EVN observed high brightness temperatures in four of our target sources supports the LLAGN origin, regardless of the exact emission mechanism. 

The best evidence for LLAGN activity however exists for the target source that has a very steep spectrum and by far the lowest EVN/WSRT flux density ratio: NGC~1272 was shown to have a pair of bent, extended radio jets on arcsecond scales \citep{McBrideMcCourt14}.  

We inspected if there are signs of AGN activity in the X-ray (\textit {Chandra} or \textit {ROSAT}), 
optical \citep[Sloan Digital Sky Survey (\textit {SDSS});][]{Alam+15} and infrared \citep[Wide-field Infrared Survey Explorer (\textit {WISE});][]{Wright+10}. To date, we believe that the strongest evidence for LLAGN activity in the five sources comes from the radio.

Among the five sources, NGC~1277, NGC~1278 and VZw~339 have been detected by \textit {Chandra} in the energy range of 0.5~--~7 keV. Their absorption-corrected luminosities, listed in Table~\ref{table2}, are on the orders of $10^{39-40}$~erg\,s$^{-1}$ \citep{Santra+07}, which is typical for LLAGNs. The three sources have a power law X-ray spectra with a photon index consisting with the expectation from LLAGNs instead of low mass X-ray binaries \citep[too steep,][]{Santra+07}. 

We also computed the ratio between the radio and X-ray luminosities for these three sources. The ratio is about $3\times10^{-4}$ for NGC~1277, $9\times10^{-3}$ for NGC~1278, and $7\times10^{-3}$ for VZw~339. They are at least one order higher than the G\"udel-Benz relation, $L_\mathrm{R}/L_\mathrm{X} \sim 10^{-5}$, which is observed by \citet{LaorBehar08} for the radio-quiet Palomar-Green quasar sample and where $L_\mathrm{R}$ is the radio luminosity at 5~GHz and $L_\mathrm{X}$ is the bolometric 0.2~--~20~keV X-ray luminosity. Together with their high brightness temperature and the absence of large variability, we are in favour of synchrotron emission from compact jets rather than free-free emission from accretion disk winds \citep[e.g.][]{BondiPerezTorres10} in the five LLAGNs. 

Optical spectroscopy exists for VZw~339 and NGC~1270 (\textit {SDSS}), and NGC~1277 \citep[][]{Trujillo+14}. These spectra show no prominent emission lines such as \ion{[O}{III]} $\lambda$5007. \cite{Kauffmann+03} show that typical strong AGNs have \ion{[O}{III]} luminosity $> 10^{7}~ \rm L_{\sun}$. The maximally allowed \ion{[O}{III]} luminosity in NGC~1270, NGC~1277 and VZw~339 is about $10^4~ \rm L_{\sun}$. This indicates that a strong AGN is not present in these galaxies and is consistent with a LLAGN scenario.
%Given the arcsec-scale resolution of the optical spectra, any low level [OIII] emission attributable to a LLAGN is likely to be washed out by the bright stellar component at the center of these galaxies.

All five targets were also observed as part of the H$\alpha$ survey by \cite{Sakai+12}. NGC~1270, NGC~1272 and NGC~1277 were not detected in this survey, implying $L_{\rm H\alpha}<10^{40}$~erg~s$^{-1}$. NGC~1278 and VZw~339 were found to have $L_{\rm H\alpha}=1\times10^{41}$~erg~s$^{-1}$ and $L_{\rm H\alpha}=2\times10^{40}$~erg~s$^{-1}$ respectively. In both cases the H$\alpha$ emission was observed to be extended on scales of 10~--~20~kpc, thus implying a non-nuclear origin. 

On the basis of data from the \textit {WISE} survey, we classified our targets using the mid-infrared colour-colour diagram \citep{Wright+10}. All of them are located in the normal elliptical galaxy regime, as shown in Figure 2, indicating that their emission is dominated by normal starlight. Note that the diagram is defined for luminous AGNs \citep{Stern+12}. In this context, the location of our target galaxies in this diagram is not surprising.

\begin{figure}
	\begin{center}
		\includegraphics[width=\columnwidth]{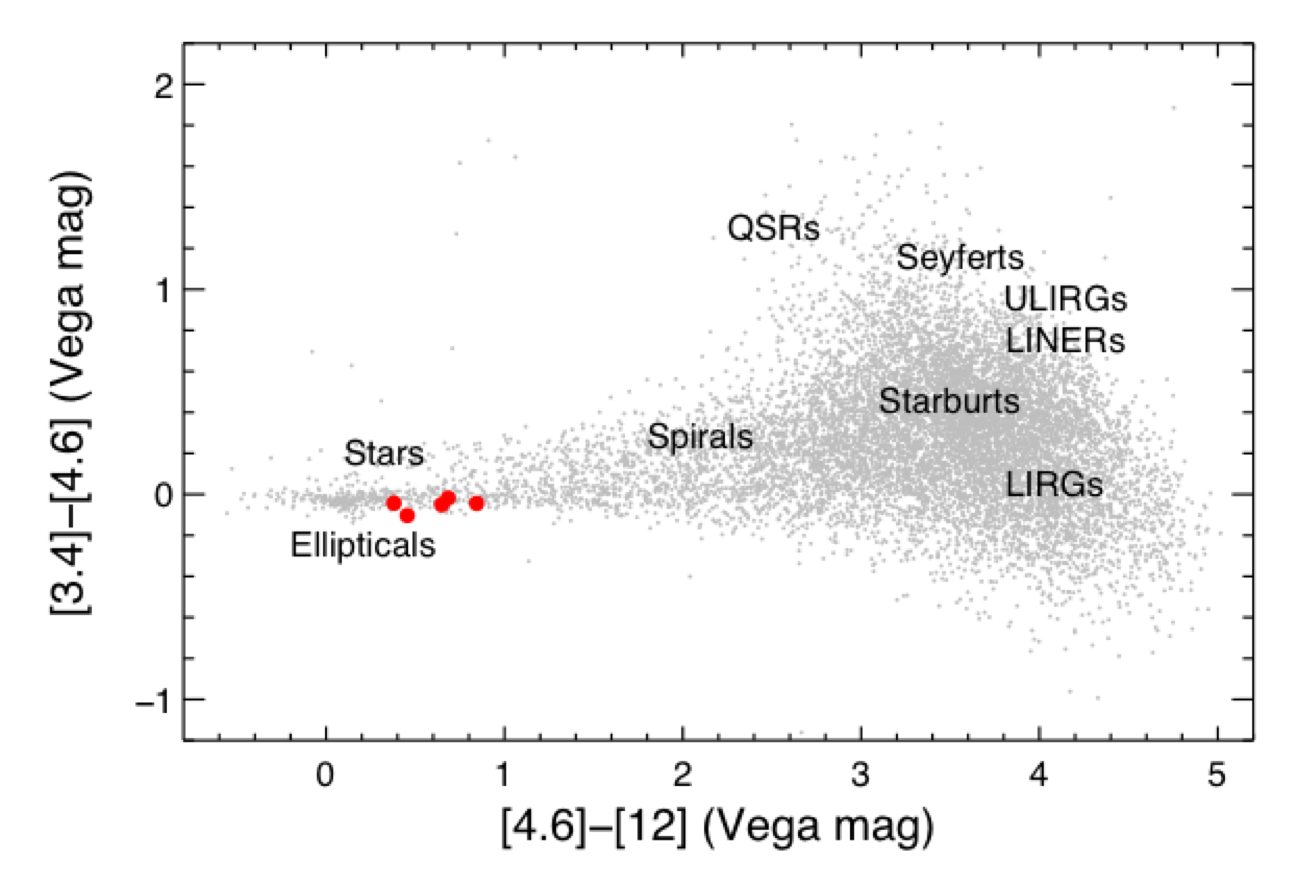}
		\label{figure2}
		\caption{Colour-colour diagram based on the \textit {WISE} infrared survey \citep[][gray dots]{Wright+10}. All of our target sources (red filled circles) are located at the normal elliptical galaxy region, which suggests that they are inactive in infrared band.}
	\end{center}
\end{figure}

\subsection{An over-massive black hole in NGC~1277?}
The compact lenticular galaxy, NGC~1277, was found by \cite{vandenBosch+12} to have
an over-massive black hole of $1.7\times10^{10}~ \rm M_{\sun}$, which corresponds to 
14 percent of the total stellar mass of that galaxy. The fraction is much larger than the typical value (0.1 percent) found in a normal massive galaxy \citep[e.g.][]{Haring+04, Sani+11}. \cite{vandenBosch+12} also noticed that this galaxy fails to follow the correlation between black hole mass and bulge luminosity. The later optical dynamical analysis by \cite{Yildirim+15} achieved a consistent mass estimation. While, recent optical studies by \cite{Graham+16} gave an order of magnitude smaller mass, $1.2\times10^{9}~ \rm M_{\sun}$. \cite{Walsh+16} found the black hole mass of $5\times10^{9}~ \rm M_{\sun}$ from near-infrared integral field unit observations, which is consistent with the result from dynamical realisations \citep{Emsellem+13}. In addition, \cite{Scharwachter2016} estimated the black hole mass of $5-17\times10^{9}~ \rm M_{\sun}$ from gas kinematics. Using infrared \textit{K}-band luminosity only, \citet{Santra+07} determined a much smaller black hole mass, $5.7\times10^{8}~ \rm M_{\sun}$. 

The radio luminosity from partially synchrotron self-absorbed compact jets 
scales with the hard X-ray luminosity in 2~--~10~keV and the mass from stellar-mass to supermassive black holes in a predictable way, i.e. the Fundamental Plane (FP) relation \citep[e.g.][]{Merloni+03, Falcke+04}. The FP relations, as regressed for the black hole mass by 
\citet{Miller-Jones+12}, using a dataset including sub-Eddington black holes 
collected by \citet{Plotkin+12}, is the following:
\begin{equation}
	{\rm log} (M_{\rm BH}) = 
	1.638\, {\rm log} (L_{\rm R}) - 1.136\, {\rm log} (L_{\rm X}) - 6.863
	\label{FP}
\end{equation}
Currently, it is not well understood under what exact conditions and up to what extremes in luminosity and mass the FP-relations may be applied. An important caveat in estimating black hole mass directly from the measured radio and X-ray luminosities is that the various instruments probe very different linear scales 
because of the large differences in angular resolution, and the measurements are 
often not contemporaneous \citep{Paragi+14}. On the other hand, if the dominating source of emission is very compact, then both instruments probe the same source of emission.   

Using the best resolution \textit {Chandra} measurement \citep{Santra+07} and our VLBI measurement,  we find a black hole mass $M_{\rm BH} \sim 4\times10^{8}~ \rm M_{\sun}$ for NGC~1277. 
This is fully in agreement with the black hole mass estimated by 
\citet{Santra+07} from its infrared luminosity. In another two sources NGC~1278 and VZw~339 with the known X-ray luminosity \citep{Santra+07}, black hole masses are found in the range of $10^{9-10}~ \rm M_{\sun}$. This is about an order of magnitude larger than derived via their infrared luminosity \citep{Santra+07}. 

The uncertainty of the mass estimation using the FP relation is certainly not small, such as $1\sigma \sim 0.44$ dex by \citet{Miller-Jones+12}. The FP relation is not a well-established tool for accurately measuring the mass of individual supermassive black holes, rather, it describes the hard-state black hole population in a statistical way. Furthermore, it is not clear whether all the X-ray emission comes from the compact jet in NGC~1277. Thus, the large difference of 1.6 dex from the optical measurement by \citet{vandenBosch+12} is still acceptable. With the FP relation, we have no significant evidence for or against an ultra-massive black hole residing in NGC~1277.    

\section{Conclusions}
With very high resolution and sensitive EVN observations of the central 10 arcminute region of the Perseus cluster, we studied the origin of radio emission in five early-type galaxies including NGC~1277, which hosts a promising ultra-massive black hole. We detected radio emission compact on the parsec scale in all the five galaxies although they have no sign of AGN activity in the optical and infrared bands. Given that these VLBI-detected radio sources are quite compact and have high brightness temperature and relatively stable flux density on time scale of years, we argue that they are linked to jet activities, powered by the central supermassive black holes in LLAGNs, instead of nuclear star formation. There is no significant evidence for or against classifying NGC~1277 as an over-massive black hole found from the fundamental plane relation. Our studies also show that VLBI is a powerful technique of searching for LLAGNs especially in case of normal galaxies.

\section*{Acknowledgements}
We thank the referee for helpful comments and suggestions.
We are grateful to Richard Plotkin for providing the data they used for regressing the 
fundamental plane relations and Ger de Bruyn for supplying the WSRT images. The European VLBI Network is a joint facility of independent European, African, Asian, and North American radio astronomy institutes. Scientific results from data presented in this publication are derived from the following EVN project code(s): EO009. \
The WSRT is operated by ASTRON (Netherlands Institute for Radio Astronomy) with support from the 
Netherlands Foundation for Scientific Research. The research leading to these results has 
received funding from the European Commission Seventh Framework Programme (FP/2007-2013) 
under grant agreement No. 283393 (RadioNet3). AIPS is produced and maintained by the 
National Radio Astronomy Observatory, a facility of the National Science Foundation operated
under cooperative agreement by Associated Universities, Inc.

%%%%%%%%%%%%%%%%%%%%%%%%%%%%%%%%%%%%%%%%%%%%%%%%%%



%\bibliographystyle{mnras}

\begin{thebibliography}{99}
\bibitem[\protect\citeauthoryear{Alam et al.}{2015}]{Alam+15} 
Alam, S., Albareti, F.~D., Allende Prieto, C., et al.\ 2015, \apjs, 219, 12 
\bibitem[\protect\citeauthoryear{Alexandroff et al.}{2012}]{Alexandroff+12} 
Alexandroff, R., Overzier, R.~A., Paragi, Z., et al.\ 2012, \mnras, 423, 1325 
\bibitem[\protect\citeauthoryear{Bell}{2003}]{Bell+03} 
Bell, E.~F.\ 2003, \apj, 586, 794 
\bibitem[\protect\citeauthoryear{Blandford \& K\"onigl}{1979}]{BlandfordKonigl79}
Blandford, R.~D., \& K{\"o}nigl, A.\ 1979, \apj, 232, 34 
\bibitem[\protect\citeauthoryear{Bondi \& P\'erez-Torres}{2010}]{BondiPerezTorres10}
Bondi, M., \& P{\'e}rez-Torres, M.-A.\ 2010, \apjl, 714, L271 
\bibitem[\protect\citeauthoryear{Bonzini et al.}{2013}]{Bonzini+13}
Bonzini, M., Padovani, P., Mainieri, V., et al.\ 2013, \mnras, 436, 3759 
\bibitem[\protect\citeauthoryear{Brown et al.}{2011}]{Brown+11}
Brown, M.~J.~I., Jannuzi, B.~T., Floyd, D.~J.~E., \& Mould, J.~R.\ 2011, \apjl, 731, L41 
\bibitem[\protect\citeauthoryear{Cao et al.}{2014}]{Cao+14}
Cao, H.-M., Frey, S., Gurvits, L.~I., et al.\ 2014, \aap, 563, A111 
\bibitem[\protect\citeauthoryear{Condon et al.}{1991}]{Condon+91}
Condon, J.~J., Huang, Z.-P., Yin, Q.~F., \& Thuan, T.~X.\ 1991, \apj, 378, 65 
\bibitem[\protect\citeauthoryear{Croton et al.}{2006}]{Croton+06}
Croton, D.~J., Springel, V., White, S.~D.~M., et al.\ 2006, \mnras, 365, 11 
\bibitem[\protect\citeauthoryear{Emsellem}{2013}]{Emsellem+13}
Emsellem, E.\ 2013, \mnras, 433, 1862  
\bibitem[\protect\citeauthoryear{Fabian et al.}{2013}]{Fabian+13} 
Fabian, A.~C., Sanders, J.~S., Haehnelt, M., Rees, M.~J., \& Miller, J.~M.\ 2013, \mnras, 431, L38 
\bibitem[\protect\citeauthoryear{Falcke et al.}{2004}]{Falcke+04} 
Falcke, H., K{\"o}rding, E., \& Markoff, S.\ 2004, \aap, 414, 895 
\bibitem[\protect\citeauthoryear{Graham et al.}{2016}]{Graham+16} 
Graham~A.W., Durr{\'e}~M., Savorgnan~G.A.D., et al.\ 2016, \apj, 819, 43
\bibitem[\protect\citeauthoryear{Greisen}{2003}]{Greisen03} 
Greisen, E.~W.\ 2003, Information Handling in Astronomy - Historical Vistas, 285, 109 
\bibitem[\protect\citeauthoryear{H{\"a}ring \& Rix}{2004}]{Haring+04} 
H{\"a}ring, N., \& Rix, H.-W.\ 2004, \apjl, 604, L89 
\bibitem[\protect\citeauthoryear{Ho et al.}{1997}]{Ho+97}
Ho, L.~C., Filippenko, A.~V., \& Sargent, W.~L.~W.\ 1997, \apjs, 112, 315 
\bibitem[\protect\citeauthoryear{Ho}{1999}]{Ho99}
Ho, L.~C.\ 1999, \apj, 510, 631 
\bibitem[\protect\citeauthoryear{Kauffmann et al.}{2003}]{Kauffmann+03} 
Kauffmann, G., Heckman, T.~M., Tremonti, C., et al.\ 2003, \mnras, 346, 1055 
\bibitem[\protect\citeauthoryear{Keimpema et al.}{2015}]{Keimpema+15}
Keimpema, A., Kettenis, M.~M., Pogrebenko, S.~V., et al.\ 2015, Experimental Astronomy, 39, 259
\bibitem[\protect\citeauthoryear{Kennicutt}{1998}]{Kennicutt+98} 
Kennicutt, R.~C., Jr.\ 1998, \apj, 498, 541  
\bibitem[\protect\citeauthoryear{Kewley et al.}{2000}]{Kewley+00}
Kewley, L.~J., Heisler, C.~A., Dopita, M.~A., et al.\ 2000, \apj, 530, 704 
\bibitem[\protect\citeauthoryear{Khan et al.}{2015}]{Khan+15}
Khan, F.~M., Holley-Bockelmann, K., \& Berczik, P.\ 2015, \apj, 798, 103   
\bibitem[\protect\citeauthoryear{Kimball et al.}{2011}]{Kimball+11}
Kimball, A.~E., Kellermann, K.~I., Condon, J.~J., Ivezi{\'c}, {\v Z}., \& Perley, R.~A.\ 2011, \apjl, 739, L29 
\bibitem[\protect\citeauthoryear{Lal \& Ho}{2010}]{LalHo10}
Lal, D.~V., \& Ho, L.~C.\ 2010, \aj, 139, 1089 
\bibitem[\protect\citeauthoryear{Laor \& Behar}{2008}]{LaorBehar08}
Laor, A., \& Behar, E.\ 2008, \mnras, 390, 847 
\bibitem[\protect\citeauthoryear{McBride \& McCourt}{2014}]{McBrideMcCourt14}
McBride, J., \& McCourt, M.\ 2014, \mnras, 442, 838
\bibitem[\protect\citeauthoryear{Merloni, Heinz \& di Matteo}{Merloni et al.}{2003}]{Merloni+03}
Merloni, A., Heinz, S., \& di Matteo, T.\ 2003, \mnras, 345, 1057
\bibitem[\protect\citeauthoryear{Miller-Jones et al.}{2012}]{Miller-Jones+12}
Miller-Jones, J.~C.~A., Sivakoff, G.~R., Altamirano, D., et al.\ 2012, \mnras, 421, 468 
\bibitem[\protect\citeauthoryear{Nagar et al.}{2000}]{Nagar+00}
Nagar, N.~M., Falcke, H., Wilson, A.~S., \& Ho, L.~C.\ 2000, \apj, 542, 186 
\bibitem[\protect\citeauthoryear{Nagar et al.}{2002}]{Nagar+02} 
Nagar, N.~M., Falcke, H., Wilson, A.~S., \& Ulvestad, J.~S.\ 2002, \aap, 392, 53 
\bibitem[\protect\citeauthoryear{Nagar et al.}{2005}]{Nagar+05}
Nagar, N.~M., Falcke, H., \& Wilson, A.~S.\ 2005, \aap, 435, 521 
\bibitem[\protect\citeauthoryear{Nyland et al.}{2016}]{Nyland+16}
Nyland, K., Young, L.~M., Wrobel, J.~M., et al.\ 2016, \mnras, 458, 2221 
\bibitem[\protect\citeauthoryear{Padovani et al.}{2011}]{Padovani+11}
Padovani, P., Miller, N., Kellermann, K.~I., et al.\ 2011, \apj, 740, 20 
\bibitem[\protect\citeauthoryear{Paragi et al.}{2014}]{Paragi+14}
Paragi, Z., Frey, S., Kaaret, P., et al.\ 2014, \apj, 791, 2
\bibitem[\protect\citeauthoryear{Perez-Torres et al.}{2009}]{Perez-Torres+09}
P{\'e}rez-Torres, M.~A., Romero-Ca{\~n}izales, C., Alberdi, A., \& Polatidis, A.\ 2009, \aap, 507, L17 
\bibitem[\protect\citeauthoryear{Plotkin et al.}{2012}]{Plotkin+12}
Plotkin, R.~M., Markoff, S., Kelly, B.~C., K{\"o}rding, E., \& Anderson, S.~F.\ 2012, \mnras, 419, 267 
\bibitem[\protect\citeauthoryear{Ranalli et al.}{2003}]{Ranalli+03}
Ranalli, P., Comastri, A., \& Setti, G.\ 2003, \aap, 399, 39 
\bibitem[\protect\citeauthoryear{Scharw{\"a}chter et al.}{2016}]{Scharwachter2016} 
Scharw{\"a}chter, J., Combes, F., Salom{\'e}, P., Sun, M., \& Krips, M.\ 2016, \mnras, 457, 4272   
\bibitem[\protect\citeauthoryear{Sadler et al.}{1989}]{Sadler+89}
Sadler, E.~M., Jenkins, C.~R., \& Kotanyi, C.~G.\ 1989, \mnras, 240, 591   
\bibitem[\protect\citeauthoryear{Sakai, Kennicutt \& Moss}{Sakai et al.}{2012}]{Sakai+12}
Sakai, S., Kennicutt, R.~C., Jr., \& Moss, C.\ 2012, \apjs, 199, 36  
\bibitem[\protect\citeauthoryear{Sani et al.}{2011}]{Sani+11} 
Sani, E., Marconi, A., Hunt, L.~K., \& Risaliti, G.\ 2011, \mnras, 413, 1479 
\bibitem[\protect\citeauthoryear{Santra et al.}{2007}]{Santra+07}
Santra, S., Sanders, J.~S., \& Fabian, A.~C.\ 2007, \mnras, 382, 895 
\bibitem[\protect\citeauthoryear{Shepherd, Pearson \& Taylor}{1994}]{Shepherd+94}
Shepherd, M.~C., Pearson, T.~J., \& Taylor, G.~B.\ 1994, \baas, 26, 987 
\bibitem[\protect\citeauthoryear{Shields \& Bonning}{2013}]{ShieldsBonning13}
Shields, G.~A., \& Bonning, E.~W.\ 2013, \apjl, 772, L5 
\bibitem[\protect\citeauthoryear{Slee et al.}{1994}]{Slee+94}
Slee, O.~B., Sadler, E.~M., Reynolds, J.~E., \& Ekers, R.~D.\ 1994, \mnras, 269, 928 
\bibitem[\protect\citeauthoryear{Somerville et al.}{2008}]{Somerville+08}
Somerville, R.~S., Hopkins, P.~F., Cox, T.~J., Robertson, B.~E., \& Hernquist, L.\ 2008, \mnras, 391, 481
\bibitem[\protect\citeauthoryear{Sijbring}{1993}]{Sijbring93}
Sijbring~D., 1993, PhD thesis, Rijksuniversiteit Groningen
\bibitem[\protect\citeauthoryear{Stern et al}{2012}]{Stern+12}
Stern, D., Assef, R.~J., Benford, D.~J., et al.\ 2012, \apj, 753, 30  
\bibitem[\protect\citeauthoryear{Trujillo et al.}{2014}]{Trujillo+14}
Trujillo~I., Ferr\'e-Mateu~A., Balcells~M., Vazdekis~A., S\'anchez-Bl\'azquez~P., 2014, ApJ, L20
\bibitem[\protect\citeauthoryear{van den Bosch et al.}{2012}]{vandenBosch+12}
van den Bosch~R.C.E. et al., 2012, \nat, 491, 729
\bibitem[\protect\citeauthoryear{Varenius et al.}{2014}]{Varenius+14}
Varenius, E., Conway, J.~E., Mart{\'{\i}}-Vidal, I., et al.\ 2014, \aap, 566, A15 
\bibitem[\protect\citeauthoryear{Weiler et al.}{2002}]{Weiler+02}
Weiler, K.~W., Panagia, N., Montes, M.~J., \& Sramek, R.~A.\ 2002, \araa, 40, 387 
\bibitem[\protect\citeauthoryear{Walsh et al.}{2016}]{Walsh+16}
Walsh, J.~L., van den Bosch, R.~C.~E., Gebhardt, K., et al.\ 2016, \apj, 817, 2
\bibitem[\protect\citeauthoryear{Wright et al.}{2010}]{Wright+10}
Wright, E.~L., Eisenhardt, P.~R.~M., Mainzer, A.~K., et al.\ 2010, \aj, 140, 1868 
\bibitem[\protect\citeauthoryear{Wrobel \& Heeschen}{1991}]{Wrobel+91}
Wrobel, J.~M., \& Heeschen, D.~S.\ 1991, \aj, 101, 148
\bibitem[\protect\citeauthoryear{Y{\i}ld{\i}r{\i}m et al.}{2015}]{Yildirim+15} 
Y{\i}ld{\i}r{\i}m, A., van den Bosch, R.~C.~E., van de Ven, G., et al.\ 2015, \mnras, 452, 1792 
\end{thebibliography}


%%%%%%%%%%%%%%%%%%%%%%%%%%%%%%%%%%%%%%%%%%%%%%%%%%

%%%%%%%%%%%%%%%%% APPENDICES %%%%%%%%%%%%%%%%%%%%%

%\appendix

%%%%%%%%%%%%%%%%%%%%%%%%%%%%%%%%%%%%%%%%%%%%%%%%%%

% Don't change these lines
\bsp	% typesetting comment
\label{lastpage}
\end{document}